# Nonuniform switching ferromagnetic layers by spin polarized current

Roger J. Elliott,[a] Ernest M. Epshtein,[b,§] Yuri V. Gulyaev,[b] Peter E. Zilberman[b*]

[a] *University of Oxford, Department of Physics, Theoretical Physics, 1 Keble Road, Oxford OX1 3NP, United Kingdom*

[b] *Institute of Radio Engineering and Electronics of the Russian Academy of Sciences, Fryazino Branch, Vvedenskii sq. 1, Fryazino, Moscow District, 141190, Russia*

**Abstract.** The magnetic reversal by spin-polarized current of a magnetic junction consisting of two ferromagnetic layers and a nonmagnetic spacer in between is considered. Initially, the free layer is magnetized antiparallel to the pinned layer by an external magnetic field. Under current flowing, a nonequilibrium spin polarization appears in the free layer. The interaction between the injected spins and the lattice leads to instability of the antiparallel orientation at the high enough current density and to switching the free layer to a state with magnetization parallel to one in the pinned layer. If the free layer thickness and the external magnetic field strength are large enough, then a nonuniform switching is favorable, so that only a part of the free layer near the injector switches. Such a switching is accompanied with appearance of a domain wall between the switched and non-switched regions. The domain wall can oscillate around the equilibrium position with some natural frequency depending on the external magnetic field.

## 1. Introduction

There is stable interest to the ferromagnetic layer magnetization reversal by spin-polarized current in magnetic junctions. The interest is supported by a number of experimental works [1 – 4], in which such switching was observed.

If the pinned and free layers of the magnetic junction are aligned originally (before the current turning on) at arbitrary angle, then the electron spin relaxation process proceeds in two stages when the electron passes from the pinned layer to free one. In the first stage, the transverse spin component relaxes, so that the spins align parallel or antiparallel to the free layer magnetization vector. The electron distribution over spin subbands remains nonequilibrium in this stage. The equilibrium distribution achieves on the longitudinal spin relaxation length that is much greater than the mentioned transverse spin relaxation length (this is the second stage).

In accordance of those two spin-relaxation stages, two possible mechanisms can be separated of the free layer switching by a high enough spin-polarized current from the pinned layer. According to one of its [5 – 7], an electron passing from the pinned layer to free one is to adapt itself to the new quantization axis, so that the spin magnetic moment transfers from electrons to the lattice and leads to the lattice magnetization reversal at high enough current density. Another switching mechanism [8 – 10] takes place in the longitudinal relaxation stage, in which electrons are aligned parallel or antiparallel to new quantization axis as a result of the transverse relaxation. If the free layer is aligned originally antiparallel to the pinned layer by means of an external magnetic field or anisotropy field, then the antiparallel state becomes unstable at some threshold current density and a switching to the parallel state occurs.

[§] E-mail: eme253@ms.ire.rssi.ru
[*] Corresponding author. Tel.: +7-095-526-9265; fax: +7-095-203-8414; e-mail: zil@ms.ire.rssi.ru

Those two mechanisms do not exclude each other. Apparently, any of them can dominate (i.e., can have lower switching threshold) depending on the experiment conditions.

A spin-injection switching mechanism due to the s-d exchange interaction of the injected nonequilibrium electron spins with the lattice magnetization was proposed in Refs. 10, 11. Injection of polarized electrons in a ferromagnetic layer aligned antiparallel to the ferromagnetic injector leads to rising magnetic energy. At high enough injection current, a reversed state of the free layer with magnetization parallel to the injector one can be favorable in energy.

Breaking the spin equilibrium in the injection process takes place at the spin diffusion length. If that length is smaller substantially than the thickness of the ferromagnetic layer in which injection happens, the uniform switching of the layer in whole thickness can be unfavorable in energy because of increasing the Zeeman energy in external magnetic field antiparallel to the injector magnetization. The switching only a part of the layer near the injector can be more favorable. At the same time, such nonuniform switching is accompanied with appearance of a domain wall that requires some energy. Conditions of the nonuniform switching as well as some associate effects are the subject of the present work.

## 2. Fundamental equation and boundary conditions

Let us introduce a spin polarization of the conduction electrons

$$P(x) = \frac{n_+(x) - n_-(x)}{n}, \qquad (1)$$

where $n_{+(-)}(x)$ are the local partial densities of the spin-up and spin-down electrons, respectively. The total electron density $n = n_+ + n_-$ is assumed constant because of the neutrality condition.

Let us derive the equation and boundary conditions for the local nonequilibrium spin polarization $\Delta P(x) = P(x) - \overline{P}$, where $\overline{P}$ is the equilibrium value of the spin polarization.

The partial current densities $j_\pm$ of the spin-up and spin-down electrons take form

$$j_\pm(x) = e\mu_\pm n_\pm(x) E(x) - eD_\pm \frac{dn_\pm(x)}{dx}, \quad (2)$$

where $E(x)$ is a local electric field, $\mu_\pm$ are partial mobilities, $D_\pm$ are partial diffusion coefficients.

The total current density

$$j = j_+ + j_- = e[\mu_+ n_+(x) + \mu_- n_-(x)]E(x) - e\left[D_+ \frac{dn_+(x)}{dx} + D_- \frac{dn_-(x)}{dx}\right] \quad (3)$$

does not depend on $x$ at steady conditions.

Excluding $E(x)$, we obtain from (2) and (3)

$$j_\pm(x) = [\mu_+ n_+(x) + \mu_- n_-(x)]^{-1}$$
$$\times \left\{\mu_\pm n_\pm(x) j \mp e\left[D_+ \mu_- n_-(x)\frac{dn_+(x)}{dx} - D_- \mu_+ n_+(x)\frac{dn_-(x)}{dx}\right]\right\}. \quad (4)$$

The partial current densities obeys the continuity equations

$$\frac{dj_\pm(x)}{dx} = \mp \frac{e}{\tau}[n_\pm(x) - \overline{n}_\pm], \quad (5)$$

where $\tau$ is the longitudinal spin relaxation time (it assumed equal for both spin orientations), $\overline{n}_\pm$ are the equilibrium partial electron densities. Hence, we obtain a continuity equation for the spin current density $J_s(x) \equiv \frac{\hbar}{2e}[j_+(x) - j_-(x)]$:

$$\frac{dJ_s(x)}{dx} = -\frac{\hbar n}{2\tau}[P(x) - \overline{P}]. \quad (6)$$

In linear approximation in $\Delta P$, the spin current density can be presented in the following form:

$$J_s = \frac{\hbar}{2e}\left(Qj + en\tilde{v}\Delta P - en\tilde{D}\frac{d\Delta P}{dx}\right), \quad (7)$$

where $Q = \dfrac{\overline{\sigma}_+ - \overline{\sigma}_-}{\sigma}$ is the spin polarization of the electric current, $\overline{\sigma}_\pm = e\mu_\pm \overline{n}_\pm$ are partial conductivities, $\sigma = \overline{\sigma}_+ + \overline{\sigma}_-$ is total conductivity,

$$\tilde{D} = \frac{D_+ \mu_- \overline{n}_- + D_- \mu_+ \overline{n}_+}{\mu_+ \overline{n}_+ + \mu_- \overline{n}_-} \quad (8)$$

is effective spin diffusion coefficient,

$$\tilde{v} = \frac{j}{en} \cdot \frac{e^2 n^2 \mu_+ \mu_-}{\sigma^2} \qquad (9)$$

is effective drift velocity. Note that $\tilde{D}$ and $\tilde{v}$ describe spin polarization (rather than charge) diffusion and drift.

We obtain an equation for $\Delta P$ from (6) and (7):

$$\frac{d^2(\Delta P)}{dx^2} - \frac{\tilde{v}}{\tilde{D}} \frac{d(\Delta P)}{dx} - \frac{\Delta P}{l^2} = 0, \qquad (10)$$

where $l \equiv \sqrt{\tilde{D}\tau}$ is a longitudinal spin relaxation length.

In fact, always $j << j_D \equiv \frac{enl}{\tau}$ in metals, so that the drift term in (10) can be neglected.

Let the current flow through the interface between two ferromagnets (the corresponding quantities are labeled with indices 1 and 2, respectively) with magnetization vectors forming some angle $\chi$. In such a situation, usually used matching conditions [12] for spin currents and partial chemical potentials require some modification.

We start from the fact (see, e.g., [13]) that the probability of the electron spin parallel to $z$ axis to be found parallel to another axis making an angle $\chi$ with $z$ axis is $\cos^2(\chi/2)$, while the probability of the antiparallel orientation is $\sin^2(\chi/2)$; sine and cosine interchange for the spin antiparallel to $z$ axis. Therefore, the electron with spin parallel to the layer 1 magnetization vector getting into the layer 2 will align parallel to the new quantization axis with probability $\cos^2(\chi/2)$ and antiparallel to it with probability $\sin^2(\chi/2)$, while the electron with spin antiparallel to the layer 1 magnetization vector will be parallel to the new axis with probability $\sin^2(\chi/2)$ and parallel to it with probability $\cos^2(\chi/2)$.

From the above, the boundary conditions for the partial currents take form of

$$j_\pm^{(2)} = j_\pm^{(1)} \cos^2 \frac{\chi}{2} + j_\mp^{(1)} \sin^2 \frac{\chi}{2}. \qquad (11)$$

Relationship (11) describes transformation of the partial current densities under rotation of the electron spin quantization axis by an angle $\chi$ (the rotation can be due to the electron transfer from one ferromagnet to another one as well as rotation of the magnetization vector of a uniform medium under external field or any other factors). Note that the transformation conserves the total current density: $j_+^{(2)} + j_-^{(2)} = j_+^{(1)} + j_-^{(1)} = j$.

A boundary condition for spin current follows from condition (11):

$$J_s^{(2)} = J_s^{(1)} \cos\chi. \tag{12}$$

A boundary condition for derivative of the spin polarization is obtained from (7) and (12):

$$j_D^{(1)} l_1 \cos\chi \frac{d(\Delta P^{(1)})}{dx} - j_D^{(2)} l_2 \frac{d(\Delta P^{(2)})}{dx} = (Q_1 \cos\chi - Q_2) j \tag{13}$$

(the drift term proportional to $\Delta P$ is omitted here).

Let us derive boundary condition for chemical potentials. A steady nonequilibrium state of a spatially nonuniform system not far from equilibrium is determined by minimum of the entropy production [14]

$$\frac{dS}{dt} = \int \frac{1}{T} \mathbf{j} \cdot \left( \mathbf{E} - \frac{1}{e}\nabla\zeta \right) dV. \tag{14}$$

In presence of two spin channels, we have

$$\frac{dS}{dt} = \frac{dS_+}{dt} + \frac{dS_-}{dt} = \int \frac{1}{T} \left( \mathbf{j} \cdot \mathbf{E} - \frac{1}{e}(\mathbf{j}_+ \cdot \nabla\zeta_+ + \mathbf{j}_- \cdot \nabla\zeta_-) \right) dV. \tag{15}$$

The entropy production $\frac{dS}{dt}$ is a scalar both in position and spin spaces. Therefore, rotation of the quantization axis conserves this quantity and, consequently, the quantity $\mathbf{j}_+ \cdot \nabla\zeta_+ + \mathbf{j}_- \cdot \nabla\zeta_-$ or $j_+\zeta_+ + j_-\zeta_-$, so that

$$j_+^{(2)}\zeta_+^{(2)} + j_-^{(2)}\zeta_-^{(2)} = j_+^{(1)}\zeta_+^{(1)} + j_-^{(1)}\zeta_-^{(1)}. \tag{16}$$

The rotation of the quantization axis leads to a linear transformation of the chemical potentials:

$$\zeta_+^{(2)} = A\zeta_+^{(1)} + B\zeta_-^{(1)}, \quad \zeta_-^{(2)} = C\zeta_+^{(1)} + D\zeta_-^{(1)}. \tag{17}$$

Substituting (11) и (17) to (16) and equating the coefficients of similar terms, we obtain a set of equations for $A$, $B$, $C$, and $D$ coefficients, from which follows

$$A = D = \frac{\cos^2\frac{\chi}{2}}{\cos\chi}, \quad B = C = -\frac{\sin^2\frac{\chi}{2}}{\cos\chi}. \tag{18}$$

Therefore,

$$\zeta_\pm^{(2)} = \frac{1}{\cos\chi}\left( \zeta_\pm^{(1)} \cos^2\frac{\chi}{2} - \zeta_\mp^{(1)} \sin^2\frac{\chi}{2} \right). \tag{19}$$

A transformation for difference of chemical potentials is obtained from (19):

$$\zeta_+^{(2)} - \zeta_-^{(2)} = \frac{1}{\cos\chi}\left( \zeta_+^{(1)} - \zeta_-^{(1)} \right). \tag{20}$$

For degenerate electrons, the partial chemical potentials $\zeta_\pm$ are related with partial electron densities $n_\pm$ by formula

$$n_\pm = \frac{(2m)^{3/2}}{6\pi^2 \hbar^3}(\zeta_\pm \pm \Delta)^{3/2}, \qquad (21)$$

where $\Delta = \mu_B B_{exc}$ is exchange splitting of the conduction band. Using the free electron approximation, we assume the effective mass $m$ not dependent on the spin direction.

Under spin equilibrium,

$$\zeta_+ = \zeta_- = \overline{\zeta} = \frac{(6\pi^2)^{2/3}\hbar^2}{2m}\overline{n}_+^{2/3} - \Delta = \frac{(6\pi^2)^{2/3}\hbar^2}{2m}\overline{n}_-^{2/3} + \Delta. \qquad (22)$$

At small deviations from the spin equilibrium, we obtain from (22) in the linear response approximation ($|n_\pm - \overline{n}_\pm| \ll \overline{n}_\pm$)

$$\zeta_\pm = \overline{\zeta} \pm \frac{2}{3}\zeta_0(1 \pm \overline{P})^{-1/3}\Delta P, \qquad (23)$$

where $\zeta_0 = \frac{(3\pi^2 n)^{2/3}\hbar^2}{2m}$.

It follows from (23)

$$\zeta_+ - \zeta_- = \frac{2}{3}\zeta_0\left[(1+\overline{P})^{-1/3} + (1-\overline{P})^{-1/3}\right]\Delta P. \qquad (24)$$

In view of the chemical potential difference matching condition (20), we obtain a boundary condition for $\Delta P$ from (24):

$$N^{(1)}\Delta P^{(1)}\cos\chi = N^{(2)}\Delta P^{(2)}, \qquad (25)$$

where

$$N = \zeta_0\left[(1+\overline{P})^{-1/3} + (1-\overline{P})^{-1/3}\right]. \qquad (26)$$

### 3. Switching conditions

Consider a system consisting of injector 1, ferromagnetic layer 2 and nonmagnetic electrode 3. Let a domain wall of thickness $L$ in distance $W$ from the injector ($0 < W < L$) exist in layer 2. The wall thickness is assumed small in comparison with the spin diffusion length $l_2$, so that the layer 2 magnetization is piecewise constant. The magnetization direction in the layer 2 is parallel to the injector magnetization at $0 < x < W$ (the coordinate $x$ is counted from the injector) and antiparallel to that at $W < x < L$. Injection of nonequilibrium spins decreases the s-d exchange energy in range $0 < x < W$ and increases it in range $W < x < L$.

Assume the electron spins pinned in the injector layer 1, so that $l_1 = 0$ (a half-metal [15] can be such ideal injector). In the nonmagnetic electrode $\overline{P}_3 = 0, Q_3 = 0$.

Since the spin polarization is assumed to be fixed in the layer 1, we seek solution of the Eq. (10) (with neglecting drift term) only for the layers 2 and 3 where it takes form

$$\Delta P(0 < x < W) = A\cosh\frac{x}{l_2} + B\sinh\frac{x}{l_2}, \tag{27}$$

$$\Delta P(W < x < L) = C\cosh\frac{x}{l_2} + D\sinh\frac{x}{l_2}, \tag{28}$$

$$\Delta P(x > L) = E\exp(-x/l_3). \tag{29}$$

On interface 1 – 2 (at $x = 0$), condition (13) is to satisfy. In this case, it takes form

$$-j_D^{(2)} l_2 \left.\frac{d\Delta P}{dx}\right|_{x=0} = (Q_1 - Q_2)j. \tag{30}$$

On the domain wall (at $x = W$), the continuity conditions for spin currents and chemical potentials (Eqs. (13) and (25)) take form

$$\Delta P|_{x=W-0} = -\Delta P|_{x=W+0}, \tag{31}$$

$$l_2\left(\left.\frac{d\Delta P}{dx}\right|_{x=W-0} + \left.\frac{d\Delta P}{dx}\right|_{x=W+0}\right) = 2Q_2\frac{j}{j_D^{(2)}}. \tag{32}$$

On interface 2 – 3 (at $x = L$), the following conditions have to satisfy:

$$N_2 \Delta P|_{x=L-0} = N_3 \Delta P|_{x=L+0}, \tag{33}$$

$$j_D^{(2)} l_2 \left.\frac{d\Delta P}{dx}\right|_{x=L-0} - j_D^{(3)} l_3 \left.\frac{d\Delta P}{dx}\right|_{x=L+0} = Q_2 j. \tag{34}$$

By substituting (27) – (29) in (30) – (34), we obtain the integration constants:

$$A = \frac{j}{j_D^{(2)}}(\sinh\lambda + \nu\cosh\lambda)^{-1}\{(Q_1 - Q_2)(\cosh\lambda + \nu\sinh\lambda)$$
$$+ Q_2[2\cosh(\lambda - w) + 2\nu\sinh(\lambda - w) - 1]\}, \tag{35}$$

$$B = -\frac{j}{j_D^{(2)}}(Q_1 - Q_2), \tag{36}$$

$$C = \frac{j}{j_D^{(2)}}(\sinh\lambda + \nu\cosh\lambda)^{-1}\{Q_2 - [Q_1 + Q_2(2\cosh w - 1)](\cosh\lambda + \nu\sinh\lambda)\}, \tag{37}$$

$$D = \frac{j}{j_D^{(2)}}[Q_1 + Q_2(2\cosh w - 1)], \tag{38}$$

where $\lambda = L/l_2$, $w = W/l_2$, $\nu = (j_D^{(3)}/j_D^{(2)})(N_2/N_3)$.

The change in the s-d exchange energy (per unit area of the layer 2) due to the spin-polarized current along $x$ axis is

$$\Delta E_{s-d} = -\alpha\mu_B n \int_0^L M(x)\Delta P(x)dx$$
$$= -\alpha\mu_B n M_0 l_2 [A\sinh w + B(\cosh w - 1) + C(\sinh\lambda - \sinh w) + D(\cosh\lambda - \cosh w)], \quad (39)$$

where $\mu_B$ is the Bohr magneton, $\alpha$ is dimensionless constant of the s-d exchange coupling, $M(x)$ is local lattice magnetization, $M_0$ is the layer 2 saturation magnetization.

The total magnetic energy of the layer 2 (per unit area) $E_{tot}$ includes, besides $\Delta E_{s-d}$, the Zeeman energy in external magnetic field (which is assumed antiparallel to the layer 1 magnetization)

$$E_H = -\widetilde{M}_0 H l_2 (\lambda - 2w) \quad (40)$$

($\widetilde{M}_0 = M_0 + \mu_B n \overline{P}$ is the equilibrium magnetization including the free electron contribution) and the domain wall energy.

Consider a situation most favorable for the nonuniform switching, when the layer 2 thickness is large comparing with the spin diffusion length ($\lambda \gg 1$). The energy of the nonuniform state with domain wall at given current and magnetic field takes form

$$U(w,J) = -J\left\{(Q_1 - Q_2)[1 - 2\exp(-w)] - 2Q_2\exp(-2w) + Q_2(1+\nu)^{-1}\right\} + h\left(w - \frac{\lambda}{2}\right) + \Gamma. \quad (41)$$

The following dimensionless variables are used here:

$$U = \frac{E_{tot}}{8\pi\widetilde{M}_0^2 l_2}, \quad \Gamma = \frac{\gamma}{8\pi\widetilde{M}_0^2 l_2}, \quad h = \frac{H}{4\pi\widetilde{M}_0}, \quad J = \frac{\alpha\mu_B n M_0}{8\pi\widetilde{M}_0^2}\frac{j}{j_D^{(2)}}.$$

The function $U(w)$ has minimum at

$$w = \ln\left[\frac{4Q_2}{Q_1 - Q_2}\left(\sqrt{1 + \frac{4Q_2 h}{(Q_1-Q_2)^2 J}} - 1\right)^{-1}\right] \equiv w_0(J). \quad (42)$$

In non-switched state where the layer 2 magnetization is parallel to the external magnetic field and antiparallel to the layer 1 magnetization, the domain wall is absent and the magnetic energy is $U_0 = U(0) - \Gamma$. At high enough current, $U(w_0) < U_0$ condition fulfils, i.e., lower energy corresponds to the switched state with domain wall than to initial (non-switched) one without domain wall. However, this necessary condition of the nonuniform switching is not sufficient one, because the lower switched state is separated from the non-switched one by a potential barrier due to formation of the domain wall. Estimates show that the barrier is high enough. Indeed, the typical energy of domain wall is $\gamma \sim 10^{-7}$ J/cm$^2$ [16]. In experiments on the magnetization reversal by spin-polarized current [17], the area which the current flowed through was of order of $10^{-10}$ cm$^2$, so that the total energy of the domain wall under nonuniform switching was $\sim 10^{-17}$ J, i.e., $\sim 100$ eV. The probability of fluctuation creating such a wall is negligibly small.

Such a situation is illustrated with Fig. 1. There the dimensionless total energy is shown at different injection currents (marked with different colors). The pairs of the same color points at $w = 0$ and $w = \lambda$ correspond to the energy with and without the domain wall energy.

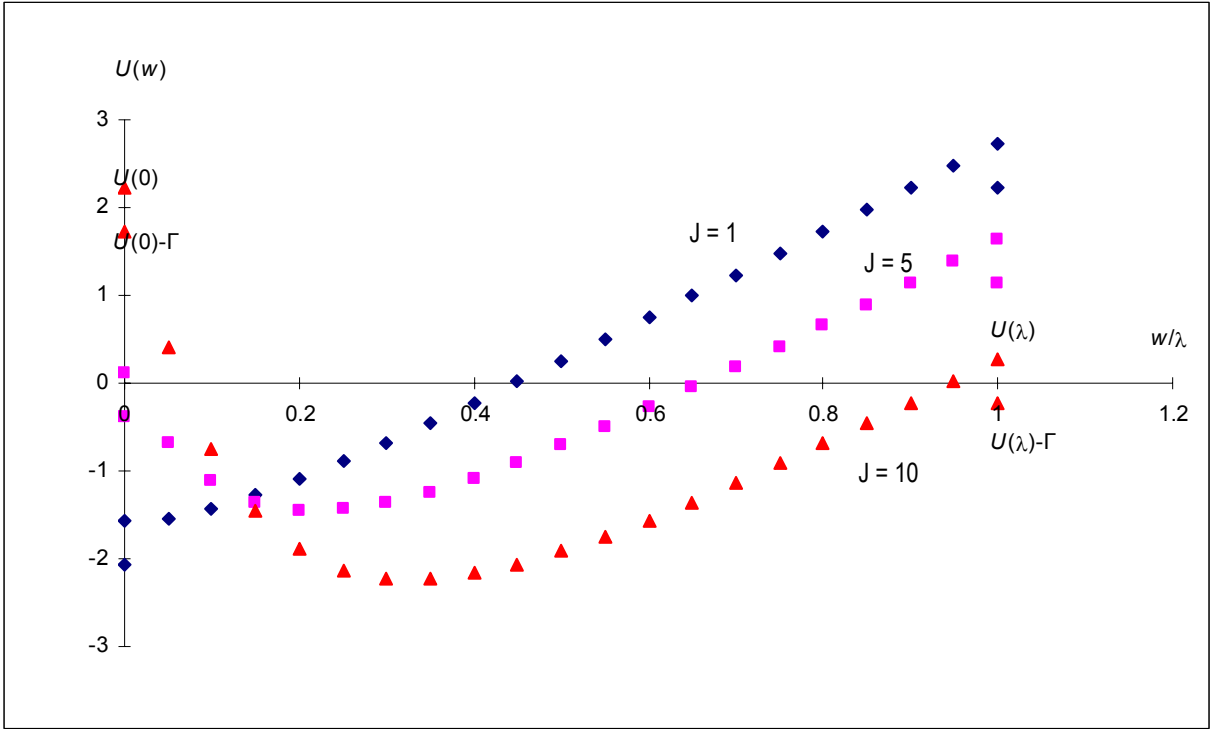

Fig. 1. The total magnetic energy at different injection currents. $Q_1 = 0.35$, $Q_2 = 0.15$, $h = 1$, $\Gamma = 0.5$, $\nu = 1$, $\lambda = 5$.

Nevertheless, nonuniform switching is possible. It was shown [10] that the state of the layer 2 with magnetization antiparallel to the injector (layer 1) magnetization becomes unstable at some threshold current density. Then the system is to go to another state. In this case, nonuniform switching occurs if the function $U(w, J)$ has a minimum below the energy of the uniformly switched state with magnetization parallel to the injector magnetization.

The dimensionless threshold current at $\lambda \gg 1$ and low magnetoresistance ratio takes form [10]

$$J_{a \to p} = \frac{(h + h_A)\lambda}{2Q_1}, \tag{43}$$

where $h_A = \frac{K}{2\pi \widetilde{M}_0^2}$, $K$ is the anisotropy constant.

In uniformly switching state ($w = \lambda$), the magnetic energy at $\lambda \gg 1$ is

$$U(\lambda, J) - \Gamma = -J\left(Q_1 - Q_2 \frac{\nu}{1+\nu}\right) + \frac{h\lambda}{2}. \tag{44}$$

The nonuniform switching condition takes form

$$U\left(w_0(J_{a \to p}), J_{a \to p}\right) < U(\lambda, J_{a \to p}) - \Gamma. \tag{45}$$

At $\lambda \gg 1$, condition (45) reduces to inequality $\lambda > \Gamma/h$. At $\Gamma \sim 10^{-7}$ J/cm$^2$, $\widetilde{M}_0 \sim 10^3$ G, $H \sim 10^2$ G that condition corresponds to the layer thickness $L \sim 50$ nm, so that it satisfies obviously at $l_2 \sim 10^{-6}$ cm and $\lambda \gg 1$.

Therefore, at the threshold current corresponding to switching [10], only part of the layer near the injector switches if the layer is thick enough. Incidentally, a domain wall appears at a distance $w_0$ from the injector (see Eq. (42)). Under further increase in the current, the domain wall shifts from the injector, i.e., the magnetization reversal is realized by means of motion of the domain wall. Under current decreasing, the domain wall shifts toward the injector and reaches it at $J = h/2(Q_1 + Q_2)$. Thus, the switching with forming domain wall displays hysteresis-type behavior. The switched part width as a function of the current is shown in Fig. 2.

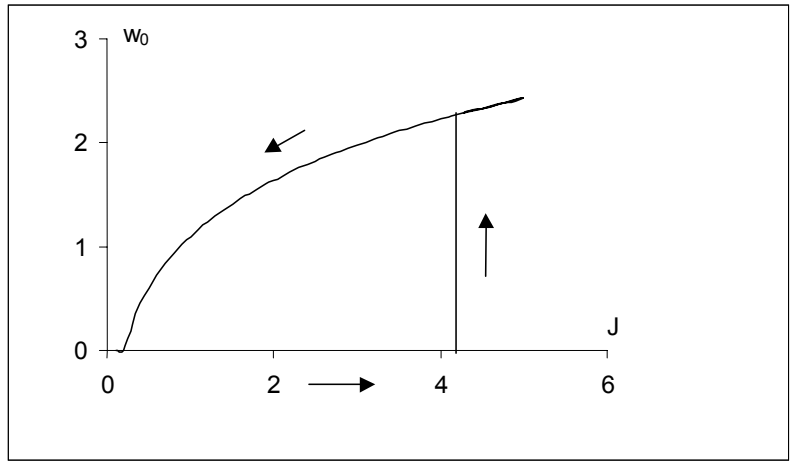

Fig. 2. The switched range width $w_0$ as a function of the injection current (in dimensionless variables). $h = 0.2$, $h_A = 0.5$, $Q_1 = 0.35$, $Q_2 = 0.15$.

## 4. Oscillation of the domain wall

When the domain wall shifts with respect to its equilibrium position corresponding to minimum of the magnetic energy, then the latter increases. It means that a restoring force affects the domain wall, so that oscillation of the domain wall with respect to the equilibrium position is possible. To estimate the oscillation frequency, we use the Döring model [16, 18], in which the domain wall motion in perfect (without pinning centers) crystal is described by following equation:

$$m\frac{d^2w}{dt^2} + \beta\frac{dw}{dt} = -\frac{8\pi\widetilde{M}^2}{l_2}\frac{dU(w)}{dw}, \quad (46)$$

where $m$ is the mass per unit area of the domain wall, $\beta$ is the damping coefficient.

The frequency of small natural oscillations of the domain wall near the equilibrium position $w = w_0$ in the potential well defined by the function $U(w)$ is

$$\omega = \sqrt{\frac{8\pi\widetilde{M}^2}{ml_2}U''(w_0)}. \quad (47)$$

Using (41) and (42), we obtain a dimensionless resonant frequency

$$\Omega \equiv \frac{\omega}{\left(\widetilde{M}/\sqrt{ml_2}\right)} = 4\sqrt{\pi J\left\{\frac{h}{J} - \frac{(Q_1-Q_2)^2}{4Q_2}\left[\sqrt{1+\frac{4Q_2 h}{(Q_1-Q_2)^2 J}}-1\right]\right\}}. \quad (48)$$

At $h = 2J(Q_1 + Q_2)$, the minimum point $w_0$ gets the origin ($w_0 = 0$), while at $h > 2J(Q_1 + Q_2)$ the function $U(w)$ has no minimum in the range $w > 0$. Therefore, the results above have a sense only under condition $h < 2J(Q_1 + Q_2)$.

At $h \ll J$, Eq. (48) simplifies considerably:

$$\Omega = \sqrt{8\pi h}. \quad (49)$$

Dependencies of the resonant frequency on the current and magnetic field are shown in Figs. 3 and 4.

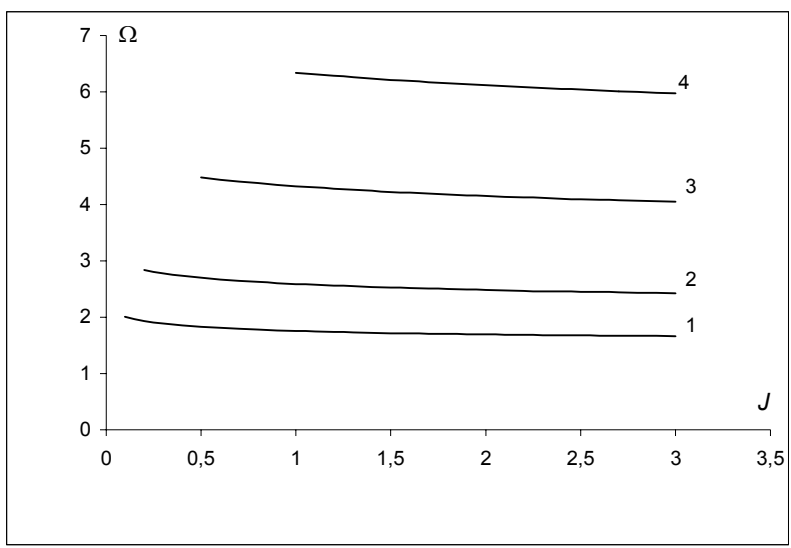

Fig. 3. Natural frequency of the domain wall oscillations as a function of the injection current (in dimensionless variables). 1 – $h = 0.1$; 2 – $h = 0.2$; 3 – $h = 0.5$; 4 – $h = 1.0$.

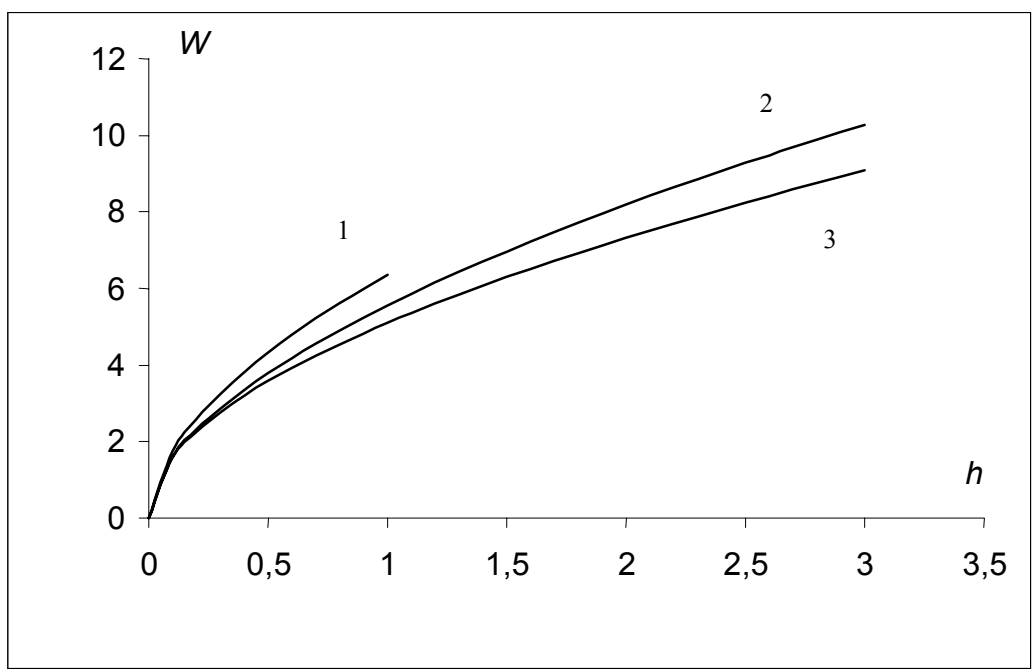

Fig. 4. Natural frequency of the domain wall oscillations as a function of the magnetic field (in dimensionless variables). 1 – $J = 1$; 2 – $J = 10$; 3 – $J = 100$.

Let us make numerical estimates. With typical values $M \sim 10^3$ G, $m \sim 10^{-10}$ g/cm$^2$, $l_2 \sim 10^{-6}$ cm we obtain $\omega \sim 10^{11}$ c$^{-1}$, so that the linear frequency of the domain wall oscillations $f =$

ω/2π is estimated by tens of GHz's. Note that the frequency ω satisfies to condition ωτ << 1; this fact justifies using the steady diffusion equation in calculations of the nonequilibrium polarization.

The work was supported by International Science & Technology Center (grant ISTC # 1522) and by Russian Foundation of Basic Research (grant RFBR # 03-02-17540).